\documentclass[%
 reprint,
 amsmath,amssymb,
 aps,
pra,
]{revtex4-2}

\usepackage{graphicx}% Include figure files
\usepackage{dcolumn}% Align table columns on decimal point
\usepackage{bm}% bold math
\usepackage{hyperref}% add hypertext capabilities
\begin{document}

\preprint{APS/123-QED}

\title{Simultaneous amplification and shaping of excimer lasers using Stimulated Brillouin Scattering in the strongly damped limit
}% Force line breaks with \\

\author{Jihoon Kim$^1$, Roopendra S Rajawat$^1$, Polina Blinova$^1$, Andrey Mironov$^2$, Milan Holec$^2$, Austin Steinforth$^2$, Conner Galloway$^2$, Jorge Rocca$^3$, Gennady Shvets$^1$}%
\affiliation{$^1$ School of Applied and Engineering Physics, Cornell University, Ithaca, NY 14850, USA. }
\affiliation{$^2$ Xcimer Energy,  Denver, CO 10325 , USA. }
\affiliation{$^3$Department of Physics, Colorado State University, CO 80523, USA. }
%\author{Austin Steinforth}%
%\author{Andrey Mironov}%
%\author{Milan Holec}%
%\author{Conner Calloway}%
%\author{Gary Eden}%
%\author{Jorge Rocca}
%\collaboration{RISE HUB collaboration}%\noaffiliation

\date{\today}% It is always \today, today,
             %  but any date may be explicitly specified

\begin{abstract}
Attaining practical Inertial Fusion Energy (IFE) depends on how efficiently one can couple the driver energy to the nuclear fusion fuel for compression and ignition. While the excimer lasers provide an efficient alternative compared to existing laser technology, it is unclear how the lasers can be harnessed to form a pulse with desired pulse shape and intensity. Stimulated Brillouin Scattering (SBS) provides a path to compressing long, energetic pulses to short intense ones. We consider the equations governing SBS in the Strongly Damped Limit (SDL) and find that it is possible to almost completely specify the final pulse shape by providing an appropriate initial seed pulse. We provide analytic expressions for reverse-engineering the initial seed shape and delineate physical limits concerning the prepulse level. 

\end{abstract}

%\keywords{Suggested keywords}%Use showkeys class option if keyword
                              %display desired
\maketitle

%\tableofcontents

%\textbf{Technological relevance of short shaped excimer laser.} What are the shit Conner is selling? One: high coupling efficiency to fuel. Two: high wall plug efficiency. Three: Gas is good. Open with Inertial Fusion.

One of key building blocks of an Inertial Fusion Energy (IFE) power plant is the \textit{driver} which compresses and simultaneously heats the fusion fuel~\cite{Zimmerman}. A high wall-plug efficiency driver that can effectively couple to the fusion target enables high-gain implosions with reasonable (few Hz) repetition rate for realization of a practical IFE plant~\cite{meier2009systems, bayramian2011compact,atzeni}. Typical laser drivers consist of a UltraViolet (UV) temporally shaped lasers with a few nanosecond duration and hundreds of Tera Watt power~\cite{Kritcher,zylstra2022burning}.  In particular, excimer lasers, due to their inherent wavelength lying in the ultraviolet(193nm for ArF, 248nm for KrF), high wall-plug efficiency (7\%), and possibility to use a gas as a medium, are promising candidates for practical IFE drivers~\cite{excimer,obenschain1996nike,okuda2000conceptual,obenschain2020direct}. 

\begin{figure}[tbh!]
    \includegraphics[width=0.5 \textwidth]{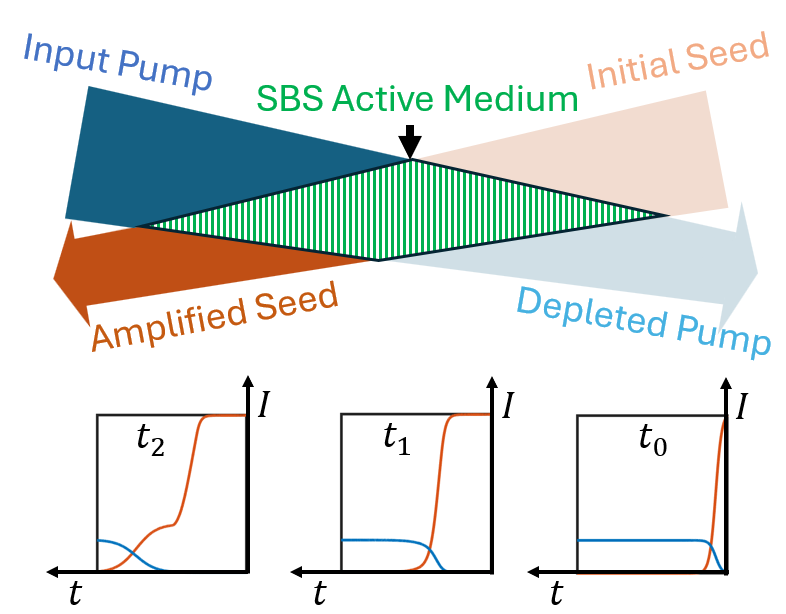}
\caption{DWG Schematics . Top: an undepleted pump (Dark Blue) and a weak seed pulse (light orange interacts in a SBS active medium (Green). The amplified seed (Dark orange) is both amplified and temporally shaped over time. Bottom: seed(orange) and pump evolution in the moving frame, progressing from right to left.  }\label{fig:schematic}
\end{figure}

However, excimer laser mediums have short excitation times, and need to be continuously pumped to generate  microseconds-long energetic pulses~\cite{excimer}. This leads to the question: how do we convert these long energetic pump pulse to a desired shaped high-power driver required for IFE? Previous works have proposed using Stimulated Brillouin Scattering (SBS)~\cite{hon,offenberger,excimer,Damzen,Montes,matsumoto} to transfer a long pump pulse energy to a counterpropagating short seed pulse in a Brillouin active medium [Fig~\ref{fig:schematic}, top row]~\cite{Pohl, Gorbunov,Gellert, Montes_Medium}. While some analytical work has been done to understand this process~\cite{Montes, Damzen, FYFChu}, an \textit{a priori} to determine the seed shape such that the final amplified pulse attain the desired temporal shape and intensity, has not been discussed in the context of SBS. 

In this letter, we describe a scheme using SBS to simultaneously compress and amplify the seed pulse, in the Strongly Damped Limit (SDL) leading to the formation of a Damped Waveform Generator (DWG).
When the seed pulse becomes sufficiently intense, a localized one-way energy transfer from the pump to seed at the seed leading edge takes place. It turns out this energy transfer in the nonlinear regime is controlled by the linear regime of the interaction, far ahead of the main body of the seed pulse. In this region, pump depletion is negligible and the amplification process can be analytically described. This leading edge controls the trailing nonlinear interaction region where pump depletion is significant, resulting in a near-analytical description of the final amplified pulse. This analytic description can be used to reverse engineer a seed pulse that would amplify to a desired final pulse shape relevant for IFE [Fig \ref{fig:schematic}, bottom row]. We  derive several analytic scalings with consequences to facility design, and delineate the limitations of this approach, showing the effect of stochastic fluctuations and the laser contrast on the amplification process.

\begin{figure}[t!]
    \includegraphics[width=0.5 \textwidth]{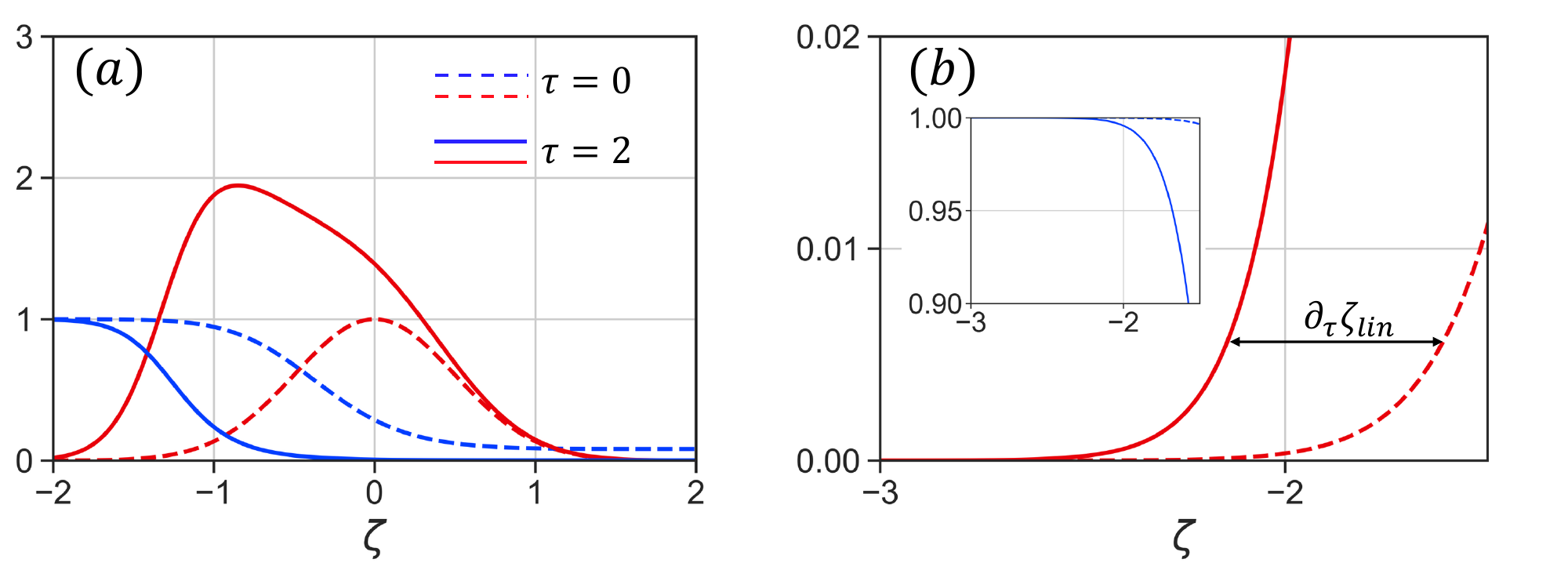}
\caption{Seed and Pump intensity evolution in the Strongly Damped limit (a-b). Seed and pump pulse intensity evolution over time (a) .  Seed(Orange) and Pump(Blue) initially ($\tau=0$, dashed) and at $\tau=2$ (solid). (b) magnificed version of (a) showing the seed pulse leading edge evolution and the seed pulse front velocity $\partial_\tau \zeta_{\rm lin}$, with inset showing pump evolution}\label{fig:evolution}
\end{figure}

\textbf{Theoretical model.}
We begin with standard normalized equation of three-wave evolution \cite{Damzen} in which a two counter-propagating seed and pump pulse interacts via a strongly damped acoustic wave: 
\begin{align}
\partial_{{z}} a + \partial_{{t}} a = -bf ,
-\partial_{{z}} b + \partial_{{t}} b = af^*,\\
 f+\frac{1}{\Gamma}\partial_t f = g ab^*\label{eq:twowave-damping}
\end{align} 
with $a$  the pump pulse amplitude, $b$ the seed pulse amplitude, $f$ the strongly dampled acoustic wave,  $g$ the gain coefficient, $\Gamma$ the damping coefficient, $z$ the propagation length and $t$ being the time~\cite{Damzen}.  See the appendix for normalizing factors. To follow the seed pulse evolution, we transform the variables from $(z,t)$ to $(\zeta,\tau)=\left( (z+t)/2,  t \right)$. Assuming phase matching (real $a,b$) and infinite damping ($\Gamma\rightarrow\infty$) , this leads to two wave equation in the strongly damped regime $
\partial_{{\zeta}} a + \partial_{{\tau}} a = -gab^2 ,
\partial_{{\tau}} b = ga^2b$. Assuming rapid pump depletion $\partial_\zeta a \gg \partial_\tau a$ leads to the following simplified equation:
\begin{align}
\partial_{{\zeta}} a = -gab^2,
\partial_{{\tau}} b = ga^2b. \label{eq:twowave}
\end{align}
The above written equations can be used for numerical and analytical investigation of the pump and seed evolution. Numerically, we assume a constant flow of pump pulse $a=a_0$ originating from $\zeta\rightarrow -\infty $ feeding the seed pulse [Fig \ref{fig:evolution} (a)]. According to the relative amplitude between the pump, seed, and the gain coefficient, the process can be classified to linear regime in which the pump depletion is negligible ($a\approx a_0$) [Fig \ref{fig:evolution} (b)], or the nonlinear regime, in which the pump depletion is significant.

In the linear regime, the seed pulse sees an undepleted pump and undergoes exponential growth. On the other hand, if the seed amplitude is appreciable, the pump amplitude is modified significantly during its interaction with the seed, resulting in slower growth. These two different types of interactions can coexist in a single smooth seed pulse at different positions [Fig \ref{fig:evolution}]. To illustrate this, we show a numerical example, in which a constant stream of pump with $a_0=1$ feeds a seed with initial amplitude $b(\tau=0, \zeta)=e^{-\zeta^2/\sigma^2}$, with $\sigma=1$. During its propagation for $\Delta \tau=2$, the gaussian seed temporal shape is skewed toward the leading edge (left) [Fig \ref{fig:evolution} (a)]. This is because the seed pulse is preferentially amplified at the leading edge ($\zeta\rightarrow -\infty$). At the leading edge, the gaussian pulse  is exponentially amplified, while the pump depletion is negligible [Fig \ref{fig:evolution} (b) ], showing clear signature of seed pulse amplification in the linear regime. 

We also note that pump sufficiently close to the intense seed peak $\zeta=0$ is almost completely depleted such that the trailing end of the seed pulse ($\zeta>0$) does not ``see" the pump nor get amplified [Fig \ref{fig:evolution}(a), orange lines]. As a result of this imbalance, the seed is amplified preferentially at the leading edge, with its front steepness increasing over propagation length .

Noting that seed pulses with sufficient intensity can completely deplete the pump pulse (Fig \ref{fig:evolution}(a), $a(\zeta\rightarrow \infty)=0$), we can establish a conservation law: 
$\partial_\tau U_b(\tau)=\partial_\tau \int_{-\infty}^{\infty}  b(\zeta',\tau)^2 d\zeta'= (2/g) \int_{-\infty}^{\infty}  a(\zeta',\tau) a_{\zeta'}(\zeta',\tau) d\zeta'= a_0^2$. This signifies , under total pump depletion, the pump energy completely being transferred to the seed pulse. Since for such high-flux seed pulses the leading front edge of the seed pulse almost completely absorbs the incoming pump energy, a measure of how fast the seed leading edge ``moves", combined with the above conservation law, can yield insight into the seed amplification process.

To track this virtual motion of the seed pulse, we define the pulse front velocity using the following condition $\frac{db}{d\tau}=\frac{\partial b}{\partial \tau}+\frac{\partial \zeta}{\partial\tau} \frac{\partial b}{\partial \zeta}=0 \rightarrow
    \frac{\partial \zeta }{\partial \tau}=-\frac{\partial_\tau b}{\partial_\zeta b}=-\frac{ga^2b}{\partial_\zeta b}
$ [Fig \ref{fig:evolution} (b)]. We consider only the front portion of a monotonically rising pulse, (i.e. $\partial_\zeta b>0 $); the seed pulse leading edge will always move ``forward" towards negative values at this leading edge( $\partial_\tau \zeta <0$). Depending on the temporal variation of the seed and pump intensity distribution, different parts of the seed  moves with different velocity. For the linear part, where the pump amplitude is constant, $a\approx a_0$, for smooth (no singularities in $\partial_\zeta b$) leading edge of pulses,  $\partial_\tau b\approx ga_0^2 b$ holds, such that $b_{\rm lin}(\tau, \zeta)\approx b_0 \exp{(ga_0^2\tau)}$, with $b_0=b(\tau=0, \zeta)$, leading to an analytically defined pulse front `velocity':
\begin{equation}
        \frac{\partial \zeta_{\rm lin} }{\partial \tau}= - \frac{ga_0^2b}{\partial_\zeta b}\approx-\frac{ga_0^2b_0}{\partial_\zeta b_0}\label{eq:length}
\end{equation}
, which is simply inverse of initial normalized seed slope. All the pump energy that is fed to the trailing part of the seed pulse is affected by  $\partial_\tau \zeta_{lin}$. Furthermore, since $b_0$ is only a function of $\zeta$, this also enables complete specification of $\partial_\tau \zeta_{lin}$ as a function of initial seed profile. 

 For concreteness, we consider several different initial pulse profiles. First, we consider a gaussian $b_0=\exp{(-\zeta^2/\sigma_0^2)}$ and a supergaussian $b_0=\exp{(-\zeta^4/\sigma_0^4)}$, with $\sigma_0=1$. According to the prescription, the linear edge of the pulse ($\zeta<0, |\zeta|\gg 1$) will move with $\partial_\tau \zeta_{\rm lin}^{\rm g}=g a_0^2\sigma_0^2/2\zeta$ for a gaussian and $\partial_\tau \zeta_{\rm lin}^{\rm sg}= ga_0^2\sigma_0^4/4\zeta^3$ for a supergaussian. This shows that the gaussian and supergaussian pulse front edge `decelerates' throughout the amplification as its position, $\zeta < 0$, decreases. We also note that the different constants $a_0, g, \sigma_0 $ defining the pump intensity, material gain, and seed shape, respectively, behave as factors regulating the pulse linear front velocity.

By combining this information on pulse front's velocity with the conservation law $\partial_\tau U_b=a_0^2$, we can deduce an asymptotic prescription for the nonlinear regime (complete pump depletion). Since this energy transfer from pump to seed occurs over a seed front propagation of $\partial_\tau \zeta_{\rm lin}$, this leads to the following prescription (see appendix \ref{sec:prescriptionproof} for proof) of pulse shape at long seed propagation ($\tau\rightarrow\infty$): 
\begin{equation}
    b^2(\zeta,\tau\rightarrow \infty)\rightarrow \frac{\partial_\tau U_b}{\partial_\tau \zeta_{\rm lin}}= -\frac{\partial_\zeta b_0}{g b_0}. \label{eq:asymptotic}
\end{equation}
 Since the rate of energy increment for the seed pulse is fixed ($\partial_\tau U_b=a_0^2$), the region over which the energy is deposited, characterized by $\partial_\tau\zeta_{\rm lin}$ and controls the local intensity to which the seed pulse gets amplified. 
 
\begin{figure}[h!]
    \includegraphics[width=0.5 \textwidth]{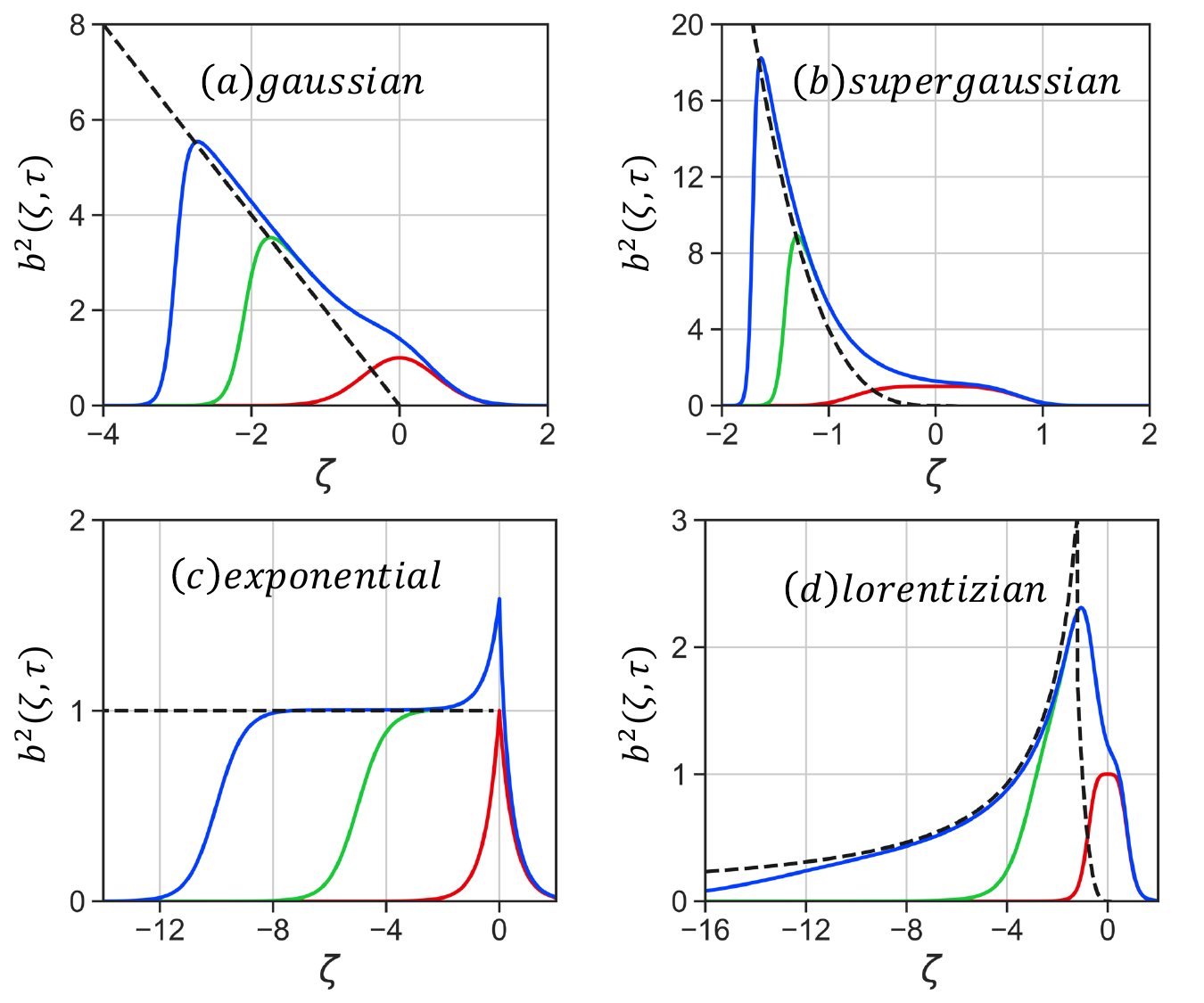}
\caption{ Asymptotic seed shape evolutions for (a)gaussian, (b) 4th order supergaussian, (c) exponential , (d) lorentzian at   $\tau=0$(solid red), $\tau=5$ (solid green) and $\tau=10$ (solid blue). Black dashed lines: analytic prescription (Eq. \ref{eq:asymptotic}). }\label{fig:asymptotics}
\end{figure}

 We return to the evolution of gaussian and supergaussian pulse shapes we described above [Fig \ref{fig:asymptotics} (a-b)] .
 
 For a gaussian pulse, integrating $\partial_\tau\zeta_{\rm lin}$ gives an asymptotic expression for the coordinate $\zeta_{\rm lin}^{\rm gauss}\rightarrow - a_0 \sigma_0 \sqrt{g \tau}$, which in turn leads to $\partial_\tau \zeta_{\rm lin}\rightarrow - \frac{ a_0\sigma_0\sqrt{g}}{2\sqrt{\tau}}$, showing that a finite amount of pump energy, $\partial_\tau U_b =a_0^2 $ is deposited to an increasingly localized region $\partial_\tau\zeta_{lin}\propto 1/\sqrt{\tau}$. This results in a peak intensity growing as $b^2(\zeta,\tau)\rightarrow -\frac{\partial_\zeta b_0}{g b_0} = - \frac{2}{\sigma_0^2 g} \zeta =  \frac{2 a_0}{\sigma_0} \sqrt{\frac{\tau}{g}} $. The combination of quadratically growing pulse length and pulse peak amplitude results in a triangular pulse shape [Fig \ref{fig:asymptotics} (a)]. 

Similar estimates can be made for supergaussian pulses, yielding $\zeta_{\rm lin}^{sg}\rightarrow - \sigma_0(ga_0^2 \tau)^{1/4}$ and peak intensity growing as $b^2(\zeta,\tau)\rightarrow  \frac{4 \sigma_0 a_0^{3/2}}{g^{1/4}}\tau^{3/4} = - \frac{4 a_0 }{g^2 \sigma_0^2} \zeta^3$. For this supergaussian pulse, the sharp pulse front slope results in virtually slower seed front velocity than that of the gaussian pulse, leading to seed amplification within a more localized region and faster peak amplitude growth rate. This results in a polynomial final seed temporal profile [Fig \ref{fig:asymptotics} (b)]. Both of these results can be checked separately by using a closed-form solutions for pump and seed pulses, which are analytically tractable in the case of supergaussian seeds and tanh-like constant pump amplitudes\cite{Japanese,FYFChu, Montes} [See Appendix].

Not all pulse shapes will have such monotonically steepening front. For instance, the pulse with exponential profile ($b_0=\exp(-|\zeta/\sigma_0|)$) has a constant leading front velocity, $\partial_\tau\zeta_{\rm lin}=- ga_0^2\sigma_0$. This leads to a formation of flat-top pulse with intensity level $b^2_{flat}=1/g\sigma_0$. The exponential amplification of linear front edge is exactly balanced by the exponential initial seed shape, resulting in a constant intensity up/down conversion at the leading edge of the seed [Fig \ref{fig:asymptotics} (c)]. The level at which it is balanced is controlled by the gain, $g$, and the slope of the exponential, $\sigma_0$, which determines the velocity seed pulse front moves and the pump photon is converted to seed photons. 

Finally, we demonstrate that more transient behaviors are also possible; a pulse with lorentzian seed ($b_0=1/(1+\zeta^4/\sigma_0^4)$), despite the waveform's resemblence to the supergaussian, undergoes a two-stage evolution. Initially, the pulse grows in intensity, but then later forms a long foot. This is due to the functional form of $\partial_\tau\zeta_{\rm lin}\propto \frac{1+\zeta^4/\sigma_0^4}{\zeta^3/\sigma_0^3}$; for $\zeta/\sigma_0\ll1$, the linear front moves as $\partial_\tau\zeta_{\rm lin}\propto \zeta^{-3}$. This leads to $\zeta_{lin}\propto \tau^{1/4}$ and $\partial_\tau\zeta_{lin}\propto \tau^{-3/4}$  leading to localized energy deposition and pulse intensity increase [Fig \ref{fig:asymptotics}(d), green]. As the linear pulse front moves, for $\zeta/\sigma_0 \gg 1$, $\partial_\tau\zeta_{\rm lin}\propto \zeta$, or $\zeta\propto \exp(C\tau)$, with $C$ a multiplicative constant. This exponential pulse length increase also signifies reduced intensity at the leading edge [Fig \ref{fig:asymptotics}(d), blue]. We also note that for this type of seed pulse lengthening, our approximation $\partial_\zeta a \gg \partial_\tau a$ which was used to derive our equations may break down for long time ($\tau\gg 1$).

We point out a few interesting points of this prescription in Eq. \ref{eq:asymptotic}.
First, the  amplified shape is only dependent on the normalized slope, not the absolute amplitude of the seed. If the normalized slope ($b_0/\partial_\zeta b_0$) of the two seed pulses are identical, the final shape to which they asymptotically grows is identical regardless of its initial amplitude. What differs is the seed pulse propagation length after which the asymptotic prescriptions hold. Second, we can prescribe piecewise slopes to "program" the pulse shape. For instance, having a piecewise exponential with different exponent coefficients connected by a gaussian slope can lead to a a flat-top pulse regions connected by a downramp [Fig \ref{fig:schematic}, bottom row, Appendix]. 

\begin{figure}[h!]
    \includegraphics[width=0.5 \textwidth]{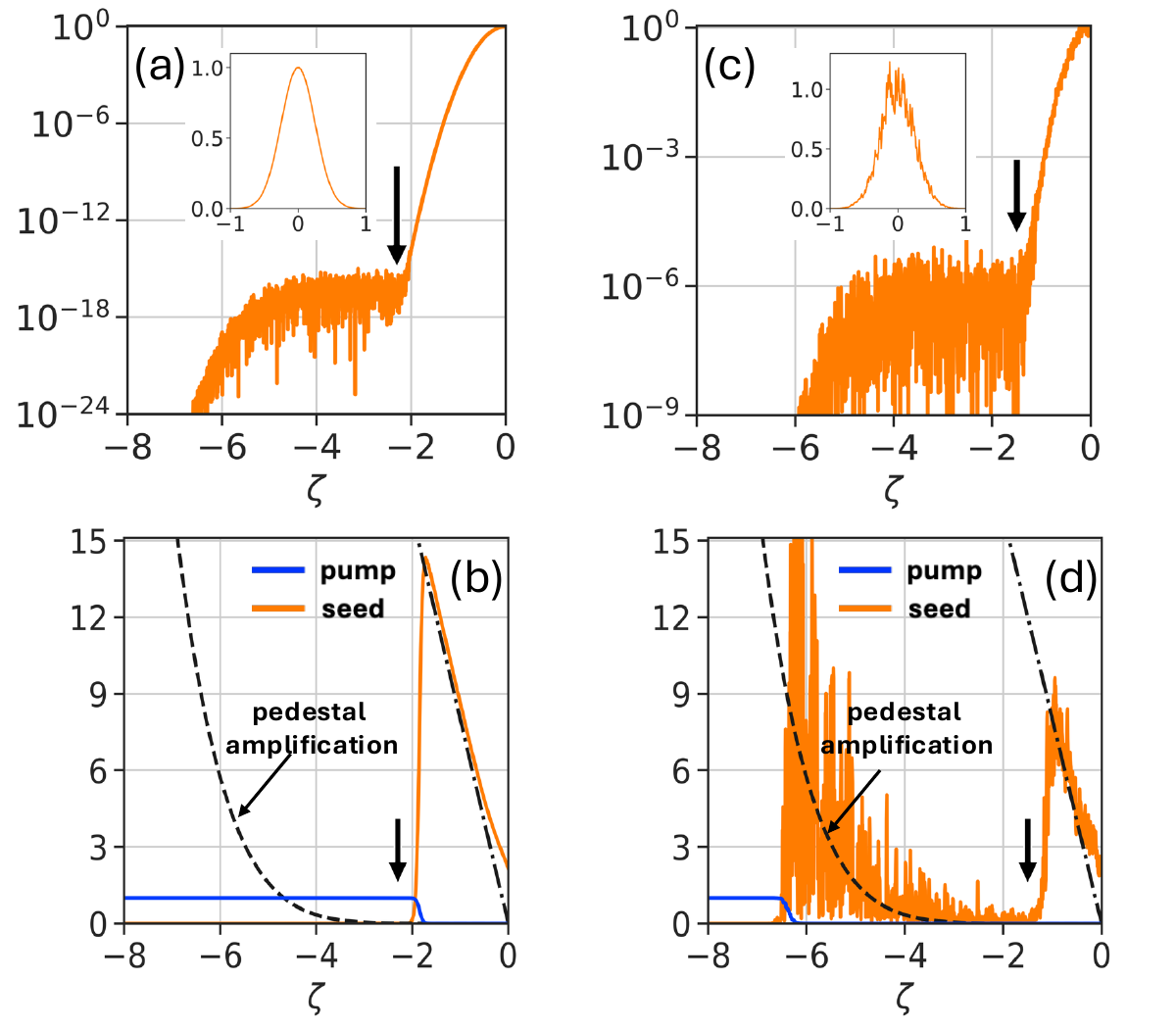}
\caption{ Effect of noise on pulse amplification. (a-b) ``clean" and (c-d) ``noisy" pulse amplification. (a,c) initial seed profiles in logarithmic scale. Insets: profile in linear scale. (b,d) seed (solid orange line) and pump (solid blue line) profile after $\tau=15$. Dash-dotted(Dashed) black lines: predicted asymptotic evolution of the main pulse(pedestal). Parameters in the main text.}\label{fig:noise}
\end{figure}

In practice, most seed pulse shape deviates from the ideal target pulse design. One way to estimate this is adding statistical noise. To estimate the effect of noise on pulse amplification, we add two types of noise: a multiplicative noise which distorts the main pulse waveform, and an additive ``noise floor" surrounding the main pulse region. Assuming a gaussian main pulse $b_0=\exp(-\zeta^2/\sigma_1^2)$, the multiplicative noise can be modeled by $b_1=b_0\times \epsilon_1 \delta_1(\zeta)$, where $\epsilon_1$ is the perturbation amplitude and $\delta_1(\zeta)$ is a gaussian random fluctuation. Similarly, the additive noise can be modeled by $b_2=\epsilon_2\exp(-\zeta^8/\sigma_2^8) \delta_2(\zeta)$, with $\delta_2$ and $\epsilon_2$ the gaussian random fluctuations and amplitude of the fluctuation, respectively. It turns out that the two types of noise exhibits distinct behavior: the multiplicative noise adds a random fluctuation on the pulse intensity, but the averaged intensity stays the same, while the additive noise sets a limit on seed contrast and pedestal dimensions. 

We illustrate this using two examples: a ``clean" [see Fig. \ref{fig:noise}(a,b)] and a ``noisy" [see Fig. \ref{fig:noise}(c,d)] seed pulse. A ``clean" seed pulse is defined by $\epsilon_1=0.001,\epsilon_2=10^{-8} $, while a ``noisy" seed pulse is defined by $\epsilon_1=0.1,\epsilon_2=10^{-3} $; the parameters defining the width of the main pulse and pedestal, $\sigma_1=0.5$ and $\sigma_2=5$ are kept identical in both cases. We first note that in the linear scale, the two seed pulses look very similar except for the amplitude modulation [see insets in Fig \ref{fig:noise} (a,c)]. However, in the logarithmic scale, the leading pedestal structure level is starkly different [see at arrow position in Fig \ref{fig:noise} (a,c)]. Because the noise structure is a wide pedestal, it crosses with the main pulse at different positions $\zeta_{trans}$ depending on the noise pedestal amplitude and the main pulse profile; in the high-contrast case, the crossover exists at $\zeta_{trans}\approx -1.5$ while in the low-contrast case, the crossover turns out to be at $\zeta_{trans}\approx -2.3$. 

This leads to the pedestal ``overwhelming" the main body of the pulse at different propagation length, $\tau$; as the seed pulse front is amplified and moves toward $\zeta\rightarrow-\infty$, the pedestal, which is in front of the main body of the pulse, would leech the main pulse energy, and eventually becomes the dominant component. For high-contrast pulses, this happens at a later time [Fig \ref{fig:noise}(b)] as opposed to the low-contrast case [Fig \ref{fig:noise} (d)] in which the main amplified pulse and the amplified noise has comparable flux. 

The transition from the main pulse amplification to noise amplification estimated by the time at which the steepened amplified pulse front arrives at $\zeta_{trans}$ where the pedestal amplitude becomes larger than the prescribed ideal pulse. Using $\zeta\rightarrow \sigma_1 \sqrt{\tau}$ ($a_0 = 1, g = 1$) which can be deduced from Eq. \ref{eq:length}, results in $\tau_{term}\approx  4\zeta^2$, yielding $\tau_{trans}^{clean} \approx 21$ for the ``clean" case and $\tau_{trans}^{noisy}\approx 9$, for the ``noisy" case which are in reasonable agreement with numerical results [Fig \ref{fig:noise}(b,d)]. Since $b^2(\zeta)\rightarrow -\frac{2}{\sigma_1^2} \zeta = -8 \zeta$, this also sets the maximum intensity $b^2$ that the main pulse can attain, namely $b_{clean}^2\approx18$ as opposed to $b_{noisy}^2\approx12$. In practice, the maximum attainable peak intensity is bit lower than this estimate. Having cleaner prepulse leads to more than 50 \% increase in peak intensity and 100\% more energy in the main pulse.

Note that the amplified pedestal behaves, on average, as amplified supergaussian, such that the extent to which they grow can be modeled by assuming a supergaussian plateau with the same width and comparable intensity level [see black color dashed lines in Fig \ref{fig:noise}(b,d)] and given as $b^2 (\zeta) \rightarrow - 8 \zeta^7/(g \sigma_2^8)$.
On the other hand, the fractional noise modulating the main body of the pulse stays as a modulation even after amplification, and the amplified main body of the pulse, on average, does follow the analytical prescription [see dash-dot lines in Fig \ref{fig:noise}(b,d)]. We find that the fractional ratio of the amplified pulse to amplified noise is identical to that of the initial seed pulse.

In conclusion, we have described a framework for designing a Stimulated Brillouin Scattering amplifier for which an initial seed pulse shape can be reverse engineered to make a desired final amplified pulse shape. By controlling the local slope of the pulse it is possible to specify the local rate at which the pump photons are converted to seed photons, thereby controlling the seed pulse temporal intensity. The controlled pulse amplification process is limited by the seed pulse contrast, but is relatively robust to the seed pulse intensity fluctuations. Such lasers can be used as efficient drivers for Inertial Fusion Energy, as well as other applications requiring shaped pulses with high tailored intensity.

 \section*{Acknowledgments}
This work is supported by the U.S. Department of Energy (DOE), Office of Science, Fusion Energy Sciences, under Award No. DE-SC0024882: IFE-STAR, DE-SC0023237.

\appendix
\section*{Appendices}

\begin{figure}[h!]
    \includegraphics[width=0.5 \textwidth]{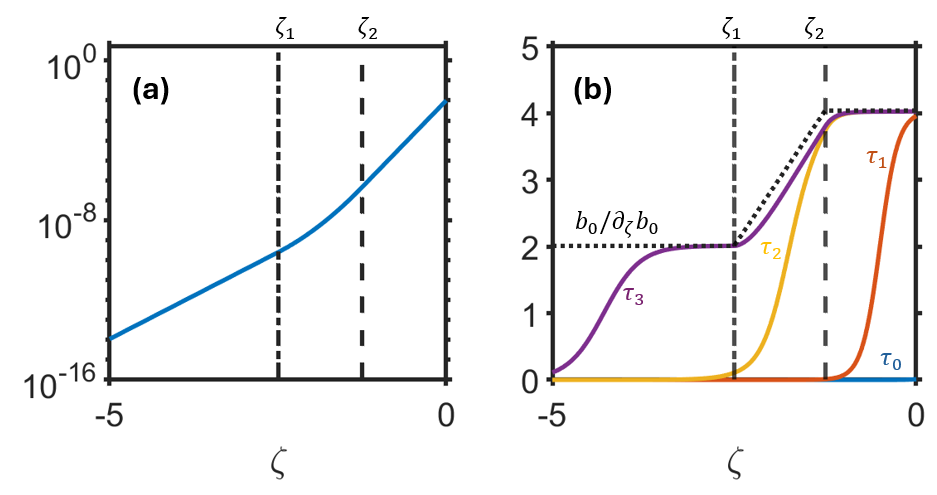}
\caption{Piecewise prescription of seed (a) and its asymptotic evolution(b). Dashed (Dash-Dotted) Lines: $\zeta_1=-2.5$, $\zeta_2=-1.25$. Dotted line in (b): asymptotic prescription. Different Colored lines in (b): amplified pulse at different propagation length $\tau_0=0$(blue), $\tau_1=5$(red), $\tau_2=10$(yellow), $\tau_3=15$(purple). }\label{fig:piecewise}
\end{figure}

\subsection{Piecewise prescription of pulse shapes.}

We show that defining initial seed pulse in a piecewise fashion can lead to complex final pulse shapes which can be relevant for IFE applications. For instance, a technologically relevant driver shape for IFE is a two step-intensity profile connected by a linear up-ramp or down-ramp in intensity that is defined by the following:

\begin{align*}
    b^2=b_1^2 \quad &\textrm{for} \quad\zeta<\zeta_1\\
    b^2=\frac{b_2^2-b_1^2}{\zeta_2-\zeta_1}(\zeta-\zeta_1)+b_1^2 \quad&\textrm{for} \quad\zeta_1<\zeta<\zeta_2\\
    b^2=b_{2}^2 \quad&\textrm{for} \quad\zeta_2<\zeta<0
\end{align*}

Using our prescription (Eq. \ref{eq:asymptotic}), such final pulse shape can be attained if the normalized pulse slope can be prescribed by $b_0=\epsilon\frac{\exp{(h(\zeta))}}{\textrm{max}(\exp{(h(\zeta)})}$, with $h(\zeta)$ the following: 

\begin{align*}
    h(\zeta)=b_1^2\zeta \quad &\textrm{for} \quad\zeta<\zeta_1\\
    h(\zeta)=\frac{b_2^2-b_1^2}{\zeta_2-\zeta_1}\frac{(\zeta-\zeta_1)^2}{2}+b_1^2 \zeta \quad&\textrm{for} \quad\zeta_1<\zeta<\zeta_2\\
    h(\zeta)=C_1+b_{2}^2(\zeta-\zeta_2) \quad&\textrm{for} \quad\zeta_2<\zeta<0\\
    \textrm{with} \quad C_1=&\frac{(b_2^2-b_1^2)(\zeta_2-\zeta_1)}{2}+b_1^2 \zeta_2
\end{align*},
and $\textrm{max}(h(\zeta))$ the maximum value of $h(\zeta)$ for $\zeta\leq 0$.

We simulate a pulse with piecewise shape given by $b_1=2, b_2=4, \zeta_1=-2.5$, $\zeta_2=-1.25$ interacting with a pump with $a_0=1$ in a medium with gain $g=1$, with  $\epsilon=0.1$. Indeed, we observe that at different times, the amplified pulse follows the piecewise asymptotic prescription. We note that there are some minor dips, which results from our solution not being exact, but rather an asymptotic `target' function.

\subsection{Comparison of selected asymptotic results with analytical formula}

The two-wave equation (Eqn.\ref{eq:twowave}) has an exact analytical solution, which can in principle be solved analytically\cite{FYFChu,Montes,Japanese}, with an analogy to migrating population of prey and predator, which in our case can be related to pump and seed. We briefly reproduce the results, and provide selected solutions to specific initial seed and pulse shapes that can be analytically solved.  

We note that letting $a^2=A, b^2=B$ and setting $g=1$ for simplicity leads to

\begin{align}
\partial_{{\zeta}} A = -AB,
\partial_{{\tau}} B = AB. \label{eq:predatorprey}
\end{align}.

We can then change the variables to 

\begin{align}
     A(\zeta,\tau)=-\partial_\tau  \log{\Delta(\zeta,\tau)}\\
     B(\zeta,\tau)=\partial_\zeta \log{\Delta(\zeta,\tau)}\\
     \Delta(\zeta,\tau)=Z(\zeta)-T(\tau)
\end{align}
with
\begin{align}
     Z(\zeta)=-\frac{1}{2} +\exp{\left(\int A d\zeta\right)}\\
     T(\tau)=\frac{1}{2}-\exp{\left(\int B d\tau\right)}
\end{align}
with the integration bounds ranging from some reasonably large range including both pulses in their frame. So if the incoming pump intensity $A$ is known for all propagation distance $\tau$ and seed distribution profile in the co-moving coordinate $\zeta$ is known, then the seed and pump behavior for all propagation length and comoving frame can be calculated. 

In practice, the integrations are only possible in closed forms for certain integrable functions, including gaussians/supergaussians and tanh functions, the former representing the seed and noise levels and the latter representing a pump that smoothly turns on/off at some finite propagation length. For example, a tanh pump and a gaussian seed pulse represented by 

\begin{align}
A(\tau)= A_0\tanh\left(\frac{\tau-\tau_p}{\sigma_p}\right)\\
B(\zeta)= B_0 \exp\left[-\left({\frac{\zeta-\zeta_b}{\sigma_b}}\right)^2\right]
\end{align}
in which $\tau_p$ is the time the pump is turned on, $\sigma_p$ is how quickly the pump is turned on, $\zeta_b$ is the center position of the gaussian, and $\sigma_b$ is the width of the seed. Assuming the seed and pump can be well-separated for a particular propagation length $\tau$, the two functions can be integrated individually, yielding the following solutions $Z(\zeta),T(\tau)$
\begin{align*}
     Z(\zeta)=-\frac{1}{2} +\exp\left[\frac{\sqrt{\pi}}{2}A_0\sigma_p\left\{\textrm{erf}{(\frac{\zeta-\zeta_p}{\sigma_p
     })}+1 \right\}\right]\\
          T(\tau)=\frac{1}{2} -\frac{\exp\left\{{-\frac{B_0}{2}(\tau-\tau_p)}\right\}}{\cosh{\left(\frac{\tau-\tau_p}{\sigma_p}\right)}^{B_0 \sigma_p/2}}\\
\end{align*}
which can be subsequently plugged in to $\Delta(\zeta,\tau)$ and be differentiated to find  the solutions for pump and seed at all position and propagation lengths.

Another relevant class of solutions for our purpose is an nth order supergaussian seed interacting with a tanh pump. 

\begin{align*}
A(\tau)= A_0\tanh\left(\frac{\tau-\tau_p}{\sigma_p}\right)\\
B(\zeta)= B_0 \exp\left\{-\left({\frac{\zeta-\zeta_b}{\sigma_b}}\right)^{2n}\right\}
\end{align*}

The solutions for these equations can still be written down as analytical functions:

\begin{align*}
     T(\tau)=\frac{1}{2} -\frac{\exp\left\{{-\frac{A_0}{2}(\tau-\tau_p)}\right\}}{\textrm{cosh}\left(\frac{\tau-\tau_p}{\sigma_p}\right)^{A_0 \sigma_p/2}}\\
          Z(\zeta)=-\frac{1}{2}+\exp{\left\{B_0 F(\zeta)\right\}}, \\
     \mathrm{If}\quad \zeta>\zeta_s,\\ F(\zeta)=-\frac{1}{2n}\zeta E_{1-\frac{1}{2n}}\left[\left(\frac{\zeta-\zeta_s}{\sigma_s}\right)^{2n}\right]+2\sigma_s \Gamma\left(1+\frac{1}{2n}\right)
     \\
    \mathrm{If} \quad\zeta<\zeta_s,\\ F(\zeta)=-\frac{1}{2n}\zeta E_{1-\frac{1}{2n}}\left[\left(\frac{\zeta-\zeta_s}{\sigma_s}\right)^{2n}\right]
    \\
E_n\equiv \int^\infty_1\frac{\exp{(-x t)}}{t^n}dt
\end{align*}
Depending on the functional form of the pump and seed, it may be possible to calculate other useful closed-form solutions.

These equations can be used to calculate, for instance, how a nth-order supergaussian pulse grows when is fed by a constant amplitude pump. We illustrate this by comparing the analytical result and numerical result for a supergaussian seed pulse defined by $B_0=1, \sigma_b=1, \zeta_b=0, n=2$ and $n=1$ being pumped by a pump defined by $A_0=1, \sigma_p=1,$ and $\tau_p=1$. This type of initial condition is approximately equal to the numerical solution of a seed with initial amplitude $b=b_0$  and duration $\sigma_0=1$ being pumped continually by $a_0=1$. Indeed, the analytical and numerical solutions show similar results, except a small difference at the front edge. This is due to the way the pump pulse is initiated in the numerical solutions; nevertheless, the body of the amplified seed pulse is almost identical for the analytical solution and the numerical solution.

\begin{figure}[h!]
    \includegraphics[width=0.5 \textwidth]{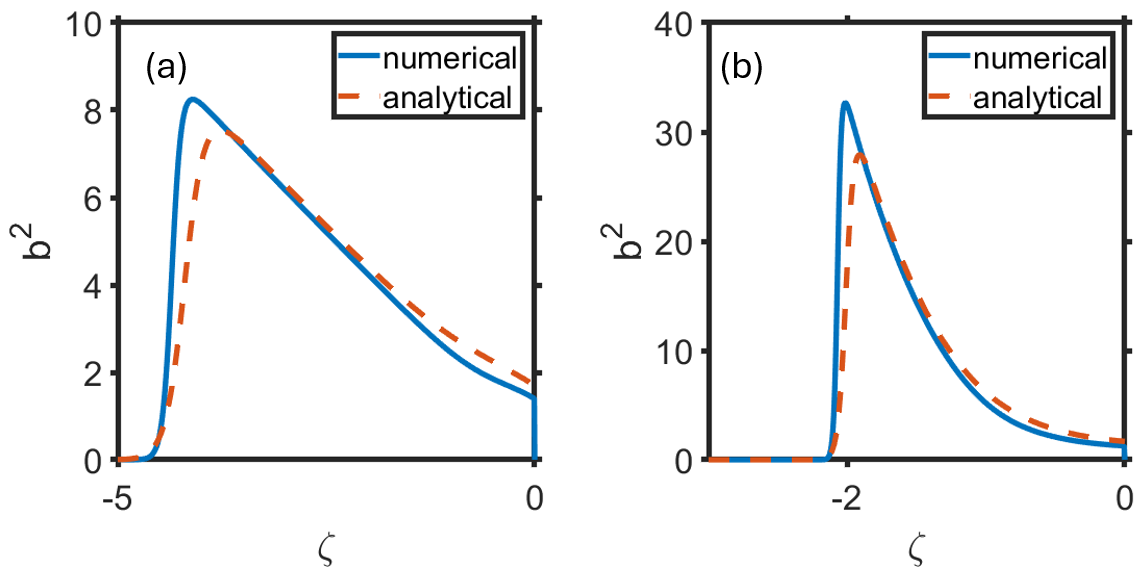}
\caption{Comparison between analytical and numerical results. (a) Gaussian seed (b) Supergaussian seed. See text for details.}\label{fig:analytics}
\end{figure}

\subsection{Equations in physical units and calculations related to Xcimer Energy's Phoenix facilities}

In the main text, we utilized normalized units. We write down the equations in physical units, and input Xcimer Energy's Phoenix laser parameters. 

In physical units, 
\begin{align}
    \textrm{Pump Depletion:} \partial_z E_a + \frac{n}{c} \partial_t E_a = i \frac{\gamma_e  \omega_a}{4nc} E_b \frac{\rho}{\rho_0}\\
    \textrm{Signal Propagation:} -\partial_z E_b + \frac{n}{c} \partial_t E_b = i \frac{\gamma_e n\omega_b}{4nc} E_a \frac{\rho^*}{\rho_0}\\
    \textrm{Acoustic Wave:} \partial_t \rho + \frac{\Gamma_b}{2} \rho = i\frac{\gamma_e \epsilon_0 K_B}{4v} E_a E_b^*
\end{align}

These are Eqn. \ref{eq:twowave-damping} but in physical units, where $E_a$ and $E_b$ are pump and signal field, respectively, $\gamma_e$ is the electrostrictive coefficient, $n$ is refractive index, $c$ being the speed of light, $v$ the acoustic wave velocity, $\omega_a$ and $\omega_b$ pump and signal frequency, $z$ propagation distance, $t$: time, $\epsilon_0$ vacuum permittivity of free space, $\Gamma_B$ Brillouin Linewidth, $\rho_0$ gas medium density, and $\rho$ density perturbation of the medium.The seed and pump intensity can be normalized to $I_{pump}=c\epsilon_0E_{p0}^2$, and the length can be normalized to wavenumber $k^{-1}$, with $k=\tilde\rho_0\gamma_e\omega/(4 n c)$. In this unit, the gain function $g=\frac{\gamma_e K_B}{2\Gamma_B c v \rho_0 \tilde{\rho}_0} I_{pump}$, with $K_B=\frac{2\pi f_B}{v}$ the acoustic wavenumber, $\rm f_B = 1.6 GHz $ the Brillouin shift and $\tilde{\rho}_0=10^{-4}$ is a small ad-hoc density perturbation scale. 

\begin{figure}[h!]
    \includegraphics[width=0.5 \textwidth]{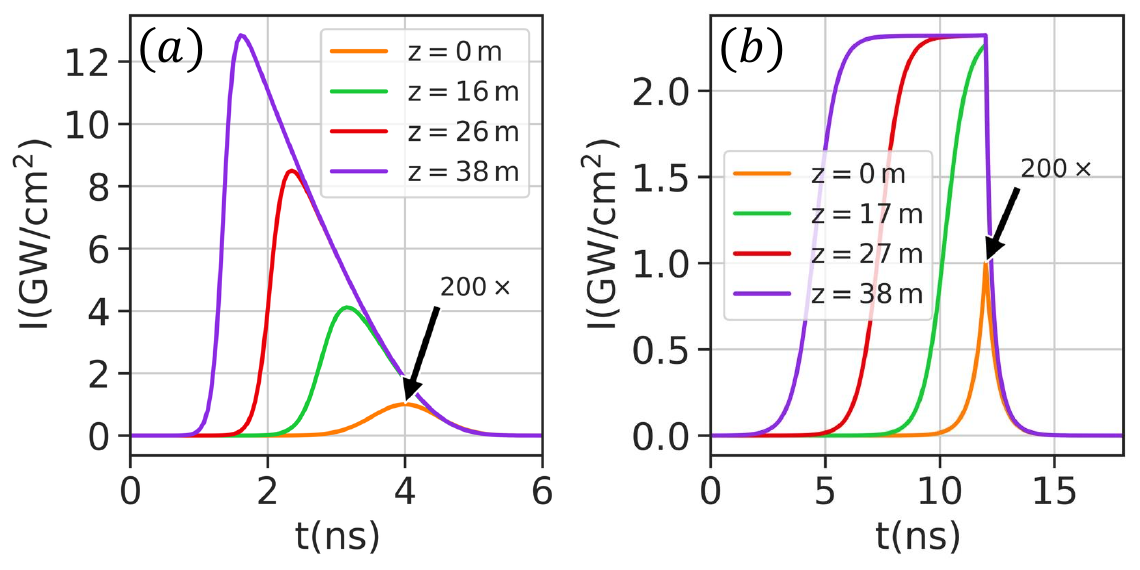}
\caption{Numerical calculation using physical units for Xcimer's Phoenix laser system for (a) gaussian and (b) exponential seed profiles. See text for parameters.}\label{fig:analytics}
\end{figure}

We set $\gamma_e\approx 2(n-1)$, and assume a Krypton gas with temperature $\rm T=273K$ and pressure $\rm P=1\textrm{atm}$. With Xcimer's Phoenix laser parameters $\rm I_{pump}=100MW/cm^2$, $ \rm I_{seed}=5MW/cm^2$, and seed duration of 1ns, this can be estimated by initial condition of a system with $ \rm g=0.49$, where a gaussian seed with $\rm \sigma_0=0.091$, $b_{peak}=0.22$ is amplified by a constant-intensity $a_0=1$ pump. Such system can be amplified to $\rm I_{seed}=13GW/cm^2$ within 38m, which is about 1000 times intensity amplification from the pump energy, compressing 260ns duration pulse into a ~1.5ns seed pulse with triangular profile for gaussian seed with 68\% extraction efficiency and into a 7.6ns seed pulse with flattop profile for an exponential seed with 60\% extraction efficiency.

\subsection{Back-of-the-envelope proof of the prescription} \label{sec:prescriptionproof}
The pulse front velocity using method of characteristics can be defined as 
\begin{equation}
\frac{db}{d\tau} = \frac{\partial b}{\partial \tau} + \frac{\partial b}{\partial \zeta}\frac{\partial \zeta}{\partial \tau} = 0,
\end{equation}
and can be written as,
\begin{align*}
\frac{\partial b^2(\zeta,\tau)}{\partial \tau} & + \frac{\partial b^2(\zeta,\tau)}{\partial \zeta}\frac{\partial \zeta}{\partial \tau} = 0, \\ 
\end{align*}
after integrating both side and some algebraic manipulation, we get
\begin{align*}
 \int_{-\infty}^{\infty} \frac{\partial b^2(\zeta,\tau)}{\partial \tau} d\zeta &+ \int_{-\infty}^{\infty} \left(\frac{\partial b^2(\zeta,\tau)}{\partial \zeta}\frac{\partial \zeta}{\partial \tau}\right) d\zeta = 0, \\
 \frac{\partial}{\partial \tau} \int_{-\infty}^{\infty} b^2(\zeta,\tau) d\zeta &+ \int_{-\infty}^{\infty} \left(\frac{\partial b^2(\zeta,\tau)}{\partial \zeta}\frac{\partial \zeta}{\partial \tau}\right) d\zeta = 0, \\
 \frac{\partial}{\partial \tau} U_b &= -\int_{-\infty}^{\infty} \left(\frac{\partial b^2(\zeta,\tau)}{\partial \zeta}\frac{\partial \zeta}{\partial \tau}\right) d\zeta.
\end{align*}
In the asymptotic limit $\tau \to \infty$, the front velocity $\partial_\tau \zeta |_{\zeta = \zeta_{lin}}$  becomes constant \cite{PolinaThesis}
\begin{align*}
\frac{\partial}{\partial \tau} U_b &= -\int_{-\infty}^{\infty} \left(\frac{\partial b^2(\zeta,\tau)}{\partial \zeta}\frac{\partial \zeta_{lin}}{\partial \tau}\right) d\zeta, \\
&= -\frac{\partial \zeta_{lin}}{\partial \tau} \int_{-\infty}^{\infty} \frac{\partial b^2(\zeta,\tau)}{\partial \zeta} d\zeta. \nonumber
\end{align*}
that yields the prescription eq. \eqref{eq:asymptotic} as
\begin{equation}
b^2_{max} = b^2(\zeta_{lin},\tau\rightarrow \infty) = \frac{\partial_\tau U_b}{\partial_\tau \zeta_{lin}}
\end{equation}
\bibliography{sbs}% Produces the bibliography via BibTeX.

@article{zylstra2022burning,
  title={Burning plasma achieved in inertial fusion},
  author={Zylstra, AB and Hurricane, OA and Callahan, DA and Kritcher, AL and Ralph, JE and Robey, HF and Ross, JS and Young, CV and Baker, KL and Casey, DT and others},
  journal={Nature},
  volume={601},
  number={7894},
  pages={542--548},
  year={2022},
  publisher={Nature Publishing Group}
}

@article{meier2009systems,
  title={Systems modeling for the laser fusion-fission energy (LIFE) power plant},
  author={Meier, WR and Abbott, R and Beach, R and Blink, J and Caird, J and Erlandson, A and Farmer, J and Halsey, W and Ladran, T and Latkowski, J and others},
  journal={Fusion science and technology},
  volume={56},
  number={2},
  pages={647--651},
  year={2009},
  publisher={Taylor \& Francis}
}

@article{obenschain2020direct,
  title={Direct drive with the argon fluoride laser as a path to high fusion gain with sub-megajoule laser energy},
  author={Obenschain, SP and Schmitt, AJ and Bates, JW and Wolford, MF and Myers, MC and McGeoch, MW and Karasik, M and Weaver, JL},
  journal={Philosophical Transactions of the Royal Society A},
  volume={378},
  number={2184},
  pages={20200031},
  year={2020},
  publisher={The Royal Society Publishing}
}

@article{bayramian2011compact,
  title={Compact, efficient laser systems required for laser inertial fusion energy},
  author={Bayramian, A and Aceves, S and Anklam, T and Baker, K and Bliss, E and Boley, C and Bullington, A and Caird, J and Chen, D and Deri, R and others},
  journal={Fusion science and Technology},
  volume={60},
  number={1},
  pages={28--48},
  year={2011},
  publisher={Taylor \& Francis}
}

@article{okuda2000conceptual,
  title={A conceptual design of a high-efficiency rep-rated pulsed power system for KrF laser drivers of inertial fusion energy},
  author={Okuda, I and Takahashi, E and Owadano, Y},
  journal={Applied Physics B},
  volume={71},
  pages={247--250},
  year={2000},
  publisher={Springer}
}

@article{obenschain1996nike,
  title={The Nike KrF laser facility: Performance and initial target experiments},
  author={Obenschain, SP and Bodner, SE and Colombant, D and Gerber, K and Lehmberg, RH and McLean, EA and Mostovych, AN and Pronko, MS and Pawley, CJ and Schmitt, AJ and others},
  journal={Physics of Plasmas},
  volume={3},
  number={5},
  pages={2098--2107},
  year={1996},
  publisher={American Institute of Physics}
}

@article{matsumoto,
  title={Efficiency and stability of pulse compression using SBS in a fiber with frequency-shifted loopback},
  author={Matsumoto, Masayuki and Miyashita, Genya},
  journal={IEEE Photonics Technology Letters},
  volume={29},
  number={1},
  pages={3--6},
  year={2016},
  publisher={IEEE}
}

@article{offenberger,
  title={Subnanosecond pulses from a KrF laser pumped SF 6 Brillouin amplifier},
  author={Fedosejevs, R and Offenberger, A},
  journal={IEEE journal of quantum electronics},
  volume={21},
  number={10},
  pages={1558--1562},
  year={1985},
  publisher={IEEE}
}

@article{hon,
  title={Pulse compression by stimulated Brillouin scattering},
  author={Hon, David T},
  journal={Optics Letters},
  volume={5},
  number={12},
  pages={516--518},
  year={1980},
  publisher={Optical Society of America}
}

@article{Zimmerman,
  title={Laser compression of matter to super-high densities: Thermonuclear (CTR) applications},
  author={Nuckolls, John and Wood, Lowell and Thiessen, Albert and Zimmerman, George},
  journal={Nature},
  volume={239},
  number={5368},
  pages={139--142},
  year={1972},
  publisher={Nature Publishing Group UK London}
}

@ARTICLE{Kritcher,
   author       = "A.L. Kritcher, et al.",
   title        = "Design of inertial fusion implosions reaching the burning plasma regime",
   journal      = "Nat. Phys.",
   volume       = "18", 
   pages        = "251-258",
   year         = "2022",
}

@article{excimer,
author = {Stephen Obenschain and Robert Lehmberg and David Kehne and Frank Hegeler and Matthew Wolford and John Sethian and James Weaver and Max Karasik},
journal = {Appl. Opt.},
number = {31},
pages = {F103--F122},
publisher = {Optica Publishing Group},
title = {High-energy krypton fluoride lasers for inertial fusion},
volume = {54},
month = {Nov},
year = {2015},
url = {https://opg.optica.org/ao/abstract.cfm?URI=ao-54-31-F103},
doi = {10.1364/AO.54.00F103},
}

@ARTICLE{Damzen,
  author={Damzen, M. and Hutchinson, H.},
  journal={IEEE Journal of Quantum Electronics}, 
  title={Laser pulse compression by stimulated Brillouin scattering in tapered waveguides}, 
  year={1983},
  volume={19},
  number={1},
  pages={7-14},
  }

@article{Montes,
  title = {Asymptotic evolution of stimulated Brillouin scattering: Implications for optical fibers},
  author = {Coste, Jean and Montes, Carlos},
  journal = {Phys. Rev. A},
  volume = {34},
  issue = {5},
  pages = {3940--3949},
  numpages = {0},
  year = {1986},
  month = {Nov},
  publisher = {American Physical Society},
  doi = {10.1103/PhysRevA.34.3940},
  url = {https://link.aps.org/doi/10.1103/PhysRevA.34.3940}
}

@article{FYFChu,
    author = {Chu, Flora Y. F. and Karney, Charles F. F.},
    title = {Solution of the three‐wave resonant equations with one wave heavily damped},
    journal = {The Physics of Fluids},
    volume = {20},
    number = {10},
    pages = {1728-1732},
    year = {1977},
    month = {10},
}

@article{Japanese,
author = {Hidenori Hasimoto},
title = {{Exact solution of a certain semi-linear system of partial differential equations related toa migrating predation problem}},
volume = {50},
journal = {Proceedings of the Japan Academy},
number = {8},
publisher = {The Japan Academy},
pages = {623 -- 627},
year = {1974},
doi = {10.3792/pja/1195518849},
URL = {https://doi.org/10.3792/pja/1195518849}
}

@article{atzeni,
author = {Stefano Atzeni},
title = {{Laser driven inertial fusion: the physical basis of current and recently proposed ignition experiments
}},
journal = {Plasma Physics and Controlled Fusion},
    volume = {51},
    number = {124029},
    year = {2009},
}

@article{Pohl,
  title = {Time-Resolved Investigations of Stimulated Brillouin Scattering in Transparent and Absorbing Media: Determination of Phonon Lifetimes},
  author = {Pohl, D. and Kaiser, W.},
  journal = {Phys. Rev. B},
  volume = {1},
  issue = {1},
  pages = {31--43},
  numpages = {0},
  year = {1970},
  month = {Jan},
  publisher = {American Physical Society},
  doi = {10.1103/PhysRevB.1.31},
  url = {https://link.aps.org/doi/10.1103/PhysRevB.1.31}
}

@article{Gorbunov,
  title = {Time compression of pulses in the course of stimulated Brillouin scattering in gases
},
  author = {Gorbunov, V A and Papernyi,  S B and Petrov,  V F and Startsev, V R},
  journal = {Soviet Journal of Quantum electronics},
  volume = {13},
  issue = {7},
  pages = {900},
  year = {1983},

}

@article{Gellert,
  title = {Investigation of stimulated brillouin scattering under well-defined interaction conditions
},
  author = {Gellert B, Kronast B},
  journal = {Appl. Phys. B},
  volume = {33},
  pages = {29-41},
  year = {1984},

}

@article{Montes_Medium,
  title = {Observation of dissipative superluminous solitons in a Brillouin fiber ring laser},
  author = {Picholle, Eric and Montes, Carlos and Leycuras, Claude and Legrand, Olivier and Botineau, Jean},
  journal = {Phys. Rev. Lett.},
  volume = {66},
  issue = {11},
  pages = {1454--1457},
  numpages = {0},
  year = {1991},
  month = {Mar},
  publisher = {American Physical Society},
  doi = {10.1103/PhysRevLett.66.1454},
  url = {https://link.aps.org/doi/10.1103/PhysRevLett.66.1454}
}

@phdthesis{PolinaThesis,
  title        = {Limits on Compression and Amplification of Excimer Lasers in Gases Using Stimulated Brillouin Scattering},
  author       = {Blinova, Polina and Shvets, Gennady},
  school       = {Cornell University},
  year         = {2025},
  type         = {thesis},
}

%\bibliographystyle{aipAuAll_links}
%\bibliographystyle{plain}

%\bibliography{sample}% Produces the bibliography via BibTeX.
\end{document}